\RequirePackage{ifpdf}
\ifpdf 
\documentclass[pdftex]{sigma}
\else
\documentclass{sigma}
\fi

\begin{document}
\allowdisplaybreaks

\renewcommand{\PaperNumber}{043}

\FirstPageHeading

\ShortArticleName{Quasigraded Lie Algebras and Modif\/ied Toda
Field Equations}

\ArticleName{Quasigraded Lie Algebras\\ and Modif\/ied Toda Field
Equations}

\Author{Taras V. SKRYPNYK~$^{\dag\ddag}$}

\AuthorNameForHeading{T.V. Skrypnyk}

\Address{$^\dag$ Bogolyubov Institute for Theoretical Physics,
       14-b  Metrologichna Str., Kyiv, 03143 Ukraine}

\Address{$^\ddag$~Institute of Mathematics, 3 Tereshchenkivs'ka
Str., Kyiv-4, 01601  Ukraine}

\EmailD{\href{mailto:tskrypnyk@imath.kiev.ua}{tskrypnyk@imath.kiev.ua}}

\ArticleDates{Received October 31, 2005, in f\/inal form March 03,
2006; Published online April 16, 2006}

\Abstract{We construct a  family of  quasigraded Lie algebras that
coincide with the defor\-mations of the loop algebras in
``principal'' gradation and admit Kostant--Adler--Symes scheme.
Using them we obtain new Volterra coupled systems and  modif\/ied
Toda f\/ield equations for all series of classical matrix Lie
algebras $\mathfrak{g}$.}


\Classification{37K05; 37K30}

\section{Introduction}
Integrability of equations of $1+1$ f\/ield theory and condensed
matter physics is based on the possibility to represent them in
the form of the zero-curvature equation \cite{ZahSh,TF,New}:
\begin{gather*}
\frac{\partial U(x,t,\lambda)}{\partial t}-\frac{\partial
V(x,t,\lambda) }{\partial x}+[U(x,t,\lambda),V(x,t,\lambda)]=0.
\end{gather*}
The most productive interpretation  of the zero-curvature equation
is achieved
 (see \cite{New,FNR,FNR1,Hl})
 if one treats it as a consistency
condition for a set of  commuting Hamiltonian f\/lows on a dual
space to some inf\/inite-dimensional Lie algebra
$\widetilde{\mathfrak{g}}$ of matrix-valued function of $\lambda$
written in the Euler--Arnold (generalized Lax) form. In this case
the corresponding $U$--$V$ pair coincides with the matrix
gradients of
 mutually commuting Hamiltonians with respect to the natural
Lie--Poisson bracket on $\widetilde{\mathfrak{g}}^*$. The method
that provides the needed set of the commuting Hamiltonian f\/lows
is the famous Kostant--Adler scheme \cite{RST3,New}. The main
ingredient of this scheme is an existence of the decomposition of
the algebra $\widetilde{\mathfrak{g}}$ into the sum of two
subalgebras: $\widetilde{\mathfrak{g}}=\widetilde{\mathfrak{g}}_+
+ \widetilde{\mathfrak{g}}_-$.

Such approach was originally  based on the graded loop algebras
$\widetilde{\mathfrak{g}}=L(\mathfrak{g})=\mathfrak{g}\otimes
P(\lambda,\lambda^{-1})$ \cite{New,FNR,FNR1,Hl} that possess
decompositions into sums of two subalgebras.  In our previous
papers \cite{Skr0,Skr1,Skr2,Skr4,Skr5} it was shown that special
quasigraded Lie algebras could be also used in the framework of
this approach. More precisely, we have constructed a family of
quasigraded Lie algebras $\widetilde{\mathfrak{g}}_A$
parameterized by some numerical matrices $A$ possessing  the
decomposition
$\widetilde{\mathfrak{g}}_A=\widetilde{\mathfrak{g}}_A^+ +
\widetilde{\mathfrak{g}}_A^-$\footnote{Subalgebras isomorphic to
$\widetilde{\mathfrak{g}}_A^-$ were independently constructed in
\cite{GS2} as a possible complementary subalgebras to the Lie
algebras of Taylor series
 in the Lie algebra of Laurent power series.}. The constructed quasigraded Lie
 algebras are {\it multiparametric deformations} of the
loop algebras: $\widetilde{\mathfrak{g}}_A$ tends to
$L(\mathfrak{g})$ in the limit $A\rightarrow 0$. They also
generalize the special elliptic $so(3)$ algebra introduced in the
papers \cite{Hol1,Hol2}.

In  our previous  papers \cite{Skr5,Skr6,Skr7,Skr7'}, using
constructed quasigraded Lie algebras $\widetilde{\mathfrak{g}}_A$
we have obtained  new hierarchies of
 integrable equations that coincide with  various generalizations
 of Landau--Lifshitz and anisotropic chiral f\/ield hierarchies.

In the present paper  we combine our previous results from
\cite{Skr4,Skr6,Skr7,Skr7'} with
 ideas of~\cite{Mikh1}
and def\/ine a new type of the quasigraded Lie algebras
$\widetilde{\mathfrak{g}}^{\rm pr}_A$ admitting
Kostant--Adler--Symes scheme and
 coinciding with  deformations of the loop algebras in the
principal gradation.
 More def\/initely, it turned out  that for a special choice of the
matrices $A$
 (that depends on the classical matrix Lie algebras
$\mathfrak{g}$) it is possible to def\/ine ``principal''
subalgebras $\widetilde{\mathfrak{g}}^{\rm
pr}_A\subset\widetilde{\mathfrak{g}}_A$
 in the  analogous way as for the case of
ordinary  loop algebras  \cite{Kac,DS}. In the case $A\rightarrow
0 $ the algebras
 $\widetilde{\mathfrak{g}}^{\rm pr}_A$ coincide with the ordinary loop
algebras in the principal gradation.

 Using the technique of \cite{FNR,FNR1,New,Hl} we develop a general scheme of obtaining
of  non-linear partial dif\/ferential equations admitting the
zero-curvature representation with the values in the Lie algebras
$\widetilde{\mathfrak{g}}^{\rm pr}_A$. For this purpose we
explicitly construct dual space, coadjoint action, its invariants
and Lie--Poisson brackets for the case of the Lie algebra
$\widetilde{\mathfrak{g}}^{\rm pr}_A$. In the result  we obtain
new integrable hierarchies satisfying zero-curvature conditions
associated   with  the Lie algebras $\widetilde{\mathfrak{g}}^{\rm
pr}_A$. We show  that among the corresponding equations there are
remarkably simple partial dif\/ferential equations which we call
``generalized Volterra coupled systems''. They have the following
form:
\begin{gather}\label{vcs0}
\partial_t u_i=u_i\beta^A_{i}(v), \qquad
\partial_x v_i= v_i \alpha_i(u),
\end{gather}
where $\beta^A_{i}$ are  linear forms (``pseudo-roots'') on the
subspace $\mathfrak{g}_{\overline{-1}}$   depending on the
``deformation'' matrix $A$ and Lie algebra $\mathfrak{g}$,
$\alpha_i \in \Pi\bigcup \{-\Theta\}$ are the linear forms (roots)
on the Cartan subalgebra
$\mathfrak{g}_{\overline{0}}=\mathfrak{h}$, $H_i$ is the basis in
$\mathfrak{h}$, $E_{-\alpha_i}$, $E_{\Theta}$ is the basis in the
subspace $\mathfrak{g}_{\overline{-1}}$ and
\[
U=u\equiv\sum\limits_{i=1}^{\dim \mathfrak{h}} u_iH_i,\qquad
V=\lambda^{-1}v\equiv\lambda^{-1} \sum\limits_{\alpha_i\in \Pi
\bigcup \Theta} v_iE_{-\alpha_i},
\]
 is the $U$--$V$-pair for the corresponding ``deformed'' zero-curvature
conditions.

It is necessary to notice that in the case $\mathfrak{g}=gl(n)$
our equations~(\ref{vcs0}) coincide with the periodic closure of
the inf\/inite Volterra coupled system considered in~\cite{ShYa}.
In the non-periodic  $gl(n)$ case it was also considered in
\cite{LesSav}. For the case of the other series of classical Lie
algebras  equations~(\ref{vcs0}) seem to be new.

Using  parametrization generalizing  one from \cite{ShYa}
equations (\ref{vcs0})could be re-written in the form of the
``modif\/ied Toda-f\/ield  equations'':
\begin{gather*}
\partial^2_{xt} \psi_i=\partial_{x} \psi_i\left(\sum\limits_{\alpha_j\in
\Pi\bigcup \{-\Theta\}}
 \beta^A_{i,j} e^{\alpha_j({\psi})}\right).
\end{gather*}

The structure of the present article is as follows: in the second
section we def\/ine the quasigraded Lie algebras
$\widetilde{\mathfrak{g}}_{A}$ and  their ``principal''
subalgebras. In the third section we def\/ine dual spaces,
Lie--Poisson brackets and Casimir functions on
$\widetilde{\mathfrak{g}}^{\rm pr}_{A}$. In the fourth section we
obtain the zero-curvature equations with the values in the twisted
subalgebras $\widetilde{\mathfrak{g}}^{\rm pr}_{A}$ and consider
the case of modif\/ied Toda f\/ield hierarchies as the main
examples they yield.

\section{ Principal grading of simple Lie algebras}\label{general}
In this subsection we will introduce necessary notations and
remind some important facts from the theory of semisimple Lie
algebras \cite{Kac,DS}.
 Let algebra
$\mathfrak{g}$ with the bracket $[\cdot ,\cdot ]$ be a simple
(reductive) classical Lie algebra of the rank $n$. Let
$\mathfrak{h}\subset \mathfrak{g}$ be its Cartan subalgebra,
$\Delta_{\pm}$ be its set of positive(negative) roots, $\Pi$ --
the set of simple roots, $H_i\in \mathfrak{h}$ the basis of Cartan
subalgebra~$E_{\alpha}$, $\alpha\in \Delta$ the corresponding root
vectors.

Let us def\/ine the so-called ``principal'' grading of
$\mathfrak{g}$ \cite{Kac}, putting:
\[
\deg H_i=0,\qquad \deg E_{\alpha_i}=1, \qquad \deg
E_{-\alpha_i}=-1.
\]
It is  evident that in such a way we obtain the grading of
$\mathfrak{g}$:
$\mathfrak{g}=\sum\limits_{k=0}^{h-1}\mathfrak{g}_{\overline{k}}$
with the graded subspaces $\mathfrak{g}_{\overline{k}}$ def\/ined
as follows: $\mathfrak{g}_{\overline{k}}=\mathrm{Span}_{C}
\{E_{\alpha}\}$ , where $\alpha $ is the root of the height $k$,
i.e.\ $\alpha=\sum\limits_{i=1}^r k_i E_{\alpha_i}$ if $\alpha\in
\Delta_+$, $\alpha=\sum\limits_{i=1}^r k_i E_{-\alpha_i}$ if
$\alpha\in \Delta_-$ and $k= \sum\limits_{i=1}^r k_i$, $h$ is the
Coxeter number of $\mathfrak{g}$. In particular
$\mathfrak{g}_{\overline{0}}=\mathfrak{h}$,
$\mathfrak{g}_{\overline{1}}=\mathrm{Span}_{C}\{E_{\alpha_i},
E_{-\theta}| \alpha_i\in \Pi\}$,
$\mathfrak{g}_{\overline{-1}}=\mathrm{Span}_{C}\{E_{-\alpha_i},
E_{\theta}| \alpha_i\in \Pi\}$ and $\theta$ is the highest root of
the height $h-1$.

Let us consider the following examples of classical matrix Lie
algebras:
\begin{example} Let  $\mathfrak{g}=gl(n)$ with the
basis $(X_{ij})_{ab}=\delta_{ia}\delta_{jb}$, $i,j\in 1,n$ and the
standard commutation relations:
\[
[X_{ij},X_{kl}]=\delta_{kj} X_{il} -\delta_{il} X_{kj}.
\]
In this case
$\mathfrak{g}_{\overline{1}}=\mathrm{Span}_{C}\{E_{\alpha_i}\equiv
X_{ii+1}, E_{-\theta}\equiv X_{n1}|i \in 1,n-1\}$,
$\mathfrak{g}_{\overline{0}}=\mathrm{Span}_{C}\{H_i\equiv
X_{ii}|i\in 1,n\}$,
$\mathfrak{g}_{\overline{-1}}=\mathrm{Span}_{C}\{E_{-\alpha_i}\equiv
X_{i+1,1}, E_{\theta}\equiv X_{1n}| i \in 1,n-1\}$ and the Coxeter
number is $h=n$.
\end{example}

\begin{example} Let
$\mathfrak{g}=so(2n+1)$, where $so(2n+1)=\{X\in gl(2n+1)|X=-s
X^\top s\}$ and $s=\mathrm{diag}(1,s_{2n})$,
$s_{2n}=\begin{pmatrix}
0 & 1_n \\
1_n &0
\end{pmatrix}$.
 In such realization the Cartan subalgebra has a basis
$H_i=X_{i+1,i+1}-X_{i+n+1,i+n+1}$ where $i=1,n$, generators of
algebra that correspond to the simple  roots are
$E_{\alpha_i}=X_{i+1,i+2}-X_{n+i+2,n+i+1}$, $i=1,n-1$,
$E_{\alpha_n}=X_{n+1,1}-X_{1,2n+1}$, their negative counterparts
are $E_{-\alpha_i}=X_{i+2,i+1}-X_{n+i+1,n+i+2}$, $i=1,n-1$,
$E_{-\alpha_n}=X_{1,1+n}-X_{2n+1,1}$. The highest root corresponds
to $E_{\theta}=X_{3 2+n}-X_{2 3+n}$, its negative counterpart
corresponds to $E_{-\theta}=X_{ 2+n 3}-X_{3+n 2}$, the Coxeter
number is $h=2n$.
\end{example}

\begin{example}Let  $\mathfrak{g}=sp(n)$,
where $sp(n)=\{X\in gl(2n)|X=w X^\top w\}$ and $w=\begin{pmatrix}
0 & 1_n \\ -1_n &0
\end{pmatrix}$.
 In such realization the Cartan subalgebra has a basis
$H_i=X_{i,i}-X_{i+n,i+n}$ where $i=1,n$, generators of algebra
that correspond to the simple  roots are
$E_{\alpha_i}=X_{i,i+1}-X_{n+i+1,n+i}$, $i=1,n-1$,
$E_{\alpha_n}=X_{n,2n}$, their negative counterparts are
$E_{-\alpha_i}=X_{i+1,i}-X_{n+i,n+i+1}$, $i=1,n-1$,
$E_{-\alpha_n}=X_{2n,n}$. The highest root corresponds to
$E_{\theta}=X_{1 1+n}$, its negative counterpart corresponds to
$E_{-\theta}=X_{ 1+n 1}$, the Coxeter number is $h=2n$.
\end{example}

\begin{example} Let  $\mathfrak{g}=so(2n)$,
where $so(2n)=\{X\in gl(2n)|X=-s X^\top s\}$ and $s\equiv s_{2n}$,
$s_{2n}=\begin{pmatrix}
0 & 1_n \\
1_n &0
\end{pmatrix}$.
 In such realization the Cartan subalgebra has a basis
$H_i=X_{i,i}-X_{i+n,i+n}$ where $i=1,n$, generators of algebra
that correspond to the simple  roots are
$E_{\alpha_i}=X_{i,i+1}-X_{n+i+1,n+i}$, $i=1,n-1$,
$E_{\alpha_n}=X_{n,2n-1}-X_{n-1,2n}$, their negative counterparts
are $E_{-\alpha_i}=X_{i+1,i}-X_{n+i,n+i+1}$, $i=1,n-1$,
$E_{-\alpha_n}=X_{2n-1,n}-X_{2n,n-1}$. The highest root
corresponds to $E_{\theta}=X_{2 1+n}-X_{1 2+n}$, its negative
counterpart corresponds to $E_{-\theta}=X_{ 1+n 2}-X_{2+n 1}$, the
Coxeter number is $h=2n-2$.
\end{example}

\section{``Principal'' quasigraded Lie algebras}

\subsection{General case} It is
well-known \cite{Kac} that having the ``principal'' grading of
$\mathfrak{g}$ it is possible  to def\/ine corresponding grading
of loop space.  Let $\mathfrak{g}=\sum\limits_{k=0}^{h-1}
\mathfrak{g}_{\overline{k}}$ be the
 $\mathbb{Z}/h \mathbb{Z}$ grading of $\mathfrak{g}$. Let us
consider the  subspace $\widetilde{\mathfrak{g}}^{\rm pr}\subset
\widetilde{\mathfrak{g}}$, where $\widetilde{\mathfrak{g}}\equiv
\mathfrak{g}\otimes P(\lambda,\lambda^{-1})$ of the following
type:
\begin{gather*}
\widetilde{\mathfrak{g}}^{\rm pr}=\bigoplus\limits_{j\in
\mathbb{Z}}\mathfrak{g}_{\overline{j}} \otimes \lambda^j.
\end{gather*}
 Here
$\overline{j}$ denotes the class of equivalence of the elements
$j\in \mathbb{Z}$ $\mathrm{mod}\  h \mathbb{Z}$. From the fact
that $[\mathfrak{g}_{\overline{i}},
\mathfrak{g}_{\overline{j}}]\subset \mathfrak{g}_{\overline{i+j}}$
it follows that $\widetilde{\mathfrak{g}}^{\rm pr}$ is a closed
Lie algebra with respect to the ordinary Lie bracket on the tensor
product:
\[
[X\otimes p(\lambda),Y\otimes q(\lambda)]=[X,Y]\otimes
p(\lambda)q(\lambda),
\]
where  $X \otimes p(\lambda), Y \otimes q(\lambda) \in
\widetilde{\mathfrak{g}}^{\rm pr}$.
 It is evident from the  def\/inition itself that
$\widetilde{\mathfrak{g}}^{\rm pr}$ is the graded Lie algebra with
the grading being def\/ined by the degrees of the spectral
parameter $\lambda$.

Let us introduce the  structure of the quasigraded Lie algebra
into the loop space $\widetilde{\mathfrak{g}}$. In order  to do
this we will   deform  Lie algebraic structure in  loop algebras
$\widetilde{\mathfrak{g}}$ in the following way
\cite{Skr4,Skr5,GS2,Hol1,Hol2,Skr6,Skr7}:
\begin{gather}\label{br0}
[X \otimes p(\lambda), Y \otimes q(\lambda)]_F=[X,Y]\otimes
p(\lambda)q(\lambda) - F(X,Y)\otimes\lambda p(\lambda) q(\lambda),
\end{gather}
 where  $X \otimes p(\lambda), Y \otimes q(\lambda) \in
\widetilde{\mathfrak{g}}$
  and the map
$F:\mathfrak{g}\times\mathfrak{g}\rightarrow \mathfrak{g}$ is skew
and satisf\/ies the following two requirements which are
equivalent to the Jacobi identities:
\begin{gather*}
{\rm (J1)} \ \ \sum\limits_{{\rm c.p.} \ \{i,j,k\}}
(F([X_i,X_j],X_k)+[F(X_i,X_j),X_k]) =0,\\
{\rm (J2)} \ \ \sum\limits_{{\rm c.p.} \
\{i,j,k\}}F(F(X_i,X_j),X_k)=0.
\end{gather*}

Now we are interested in a possibility of def\/ining  the
 structure of the quasigraded algebra  on the space
 $\widetilde{\mathfrak{g}}^{\rm pr}$. For this purpose we want
bracket (\ref{br0}) to be correctly restricted to the space
$\widetilde{\mathfrak{g}}^{\rm pr}$. By the direct verif\/ication
one can prove the following proposition:
\begin{proposition}\label{au} The subspace $\widetilde{\mathfrak{g}}^{\rm pr}\subset \widetilde{\mathfrak{g}}$ is
 the closed Lie algebra
 if and only if:
\begin{gather}\label{aut}
F(\mathfrak{g}_{\overline{i}},\mathfrak{g}_{\overline{j}})
\subset\mathfrak{g}_{\overline{i+j+1}}.
\end{gather}
\end{proposition}
In the next subsection we will  present examples of the cocycles
$F$ on the f\/inite-dimensional Lie algebras $\mathfrak{g}$ that
satisfy conditions (J1), (J2) and (\ref{aut}).

\subsection{Case of classical matrix Lie algebras}

Let us now  consider the classical matrix Lie algebras
$\mathfrak{g}$ of the type $gl(n)$, $so(n)$ and $sp(n)$ over the
f\/ield $\mathbb{K}$ of the complex or real numbers. As in the
above examples  we will realize the algebra $so(n)$ as algebra of
skew-symmetric matrices: $so(n)=\{X\in gl(n)|X=-sX^\top s \}$ and
the algebra $sp(n)$  as the following matrix Lie algebra:
$sp(n)=\{X\in gl(2n)|X=wX^\top w\}$, where $s\in {\rm symm}\,(n)$,
$s^2=1$, $w\in so(2n)$ and $w^2=-1$.

Let us consider the cochain
$F:\mathfrak{g}\times\mathfrak{g}\rightarrow \mathfrak{g}$ of the
following  form:
\[
F_A(X,Y)=[X,Y]_A\equiv XAY-YAX.
\]
 From the
theory of consistent Poisson brackets it  is known~\cite{CP} to
satisfy conditions (J1), (J2).

The following Proposition holds true:
\begin{proposition}\label{cond}
The cochain $F_A$ satisfies condition \eqref{aut}  if and only if
the
 matrix  $A$ has the  form:
\begin{alignat*}{3}
& 1)\quad && A=\sum\limits_{i=1}^{n-1}a_iX_{ii+1}+a_nX_{n1} \quad {\rm if} \quad \mathfrak{g}=gl(n);&\\
& 2) && A=\sum\limits_{i=1}^{n-1}a_i( X_{i+1,i+2}+X_{n+i+2,n+i+1})
+a_n(X_{1+n,1}+X_{1,2n+1})+a_{n+1}(X_{ 2+n 3}+X_{3+n 2}) &\\
& && \mbox{if}\quad \mathfrak{g}=so(2n+1); & \\
& 3) && A=\sum\limits_{i=1}^{n-1}a_i( X_{i,i+1}+X_{n+i+1,n+i}) \quad {\rm if}\quad \mathfrak{g}=sp(n); &\\
& 4) && A=\sum\limits_{i=1}^{n-1}a_i( X_{i,i+1}+X_{n+i+1,n+i})
+a_n(X_{n,2n-1}+X_{n-1,2n})+a_{n+1}(X_{
1+n 2}+X_{2+n 1}) &\\
&&& {\rm if} \quad \mathfrak{g}=so(2n),&
\end{alignat*}
where $X_{ij}$ is the standard matrix basis of $gl(n)$,
$(X_{ij})_{\alpha\beta}=\delta_{i\alpha}\delta_{\gamma\beta}$.
\end{proposition}

\begin{proof} For the bracket constructed with the help of the
cocycle $F_A$ to be correctly def\/ined
on~$\widetilde{\mathfrak{g}}^{\rm pr}$ we have to require the
linear space $\mathfrak{g}$ to be closed with respect to the
bracket $[\cdot ,\cdot ]_A$ and linear
space~$\widetilde{\mathfrak{g}}^{\rm pr}$  as a space of
matrix-valued function of $\lambda$ to be closed with respect to
the bracket~(\ref{br}). These conditions are equivalent to the
following requirement:  $[X, Y ]_A\in
 \mathfrak{g}_{i+j+1}$ $\forall \, X\in \mathfrak{g}_{i}$, $Y\in
\mathfrak{g}_{j}$. The straightforward   verif\/ication shows in
each case  that this requirement is satisf\/ied if and only if
matrix $A$ has the form described in the proposition.
\end{proof}

Hence in the case of the matrix Lie algebras and the matrices $A$
def\/ined in the above Proposition  we may introduce into the
space $\widetilde{\mathfrak{g}}^{\rm pr}$ the new Lie bracket of
the form:
\begin{gather}\label{br}
[X \otimes p(\lambda), Y\otimes q(\lambda)]=[X,Y]\otimes
p(\lambda)q(\lambda) -[X,Y]_A \otimes \lambda
p(\lambda)q(\lambda),
\end{gather}
 where  $X \otimes p(\lambda), Y \otimes q(\lambda) \in
\widetilde{\mathfrak{g}}^{\rm pr}$, $[X, Y]\equiv XY-YX$  in the
righthand side of this identity denote ordinary Lie bracket in
$\mathfrak{g}$ and $[X,Y]_A\equiv XAY-YAX$.

\medskip

\noindent {\bf Definition.} We will denote the linear space
$\widetilde{\mathfrak{g}}^{\rm pr}$ with the bracket given by
(\ref{br}) by $\widetilde{\mathfrak{g}}_A^{\rm pr}$.
\medskip

From the very def\/inition of $\widetilde{\mathfrak{g}}_{A}^{\rm pr}$ it
follows that the algebra $\widetilde{\mathfrak{g}}_{A}^{\rm pr}$
is $Z$-quasigraded  and $\widetilde{\mathfrak{g}}^{\rm pr}_A$  admits
the direct sum decomposition $\widetilde{\mathfrak{g}}_{A}^{\rm
pr}= \widetilde{\mathfrak{g}}_{A}^{\rm pr+}+
\widetilde{\mathfrak{g}}_{A}^{\rm pr-},$ where
\begin{gather*}
\widetilde{\mathfrak{g}}_{A}^{{\rm pr}+}=\bigoplus\limits_{j\geq 0
}\mathfrak{g}_{\overline{j}} \otimes \lambda^j,\qquad
\widetilde{\mathfrak{g}}_{A}^{{\rm
pr}-}=\bigoplus\limits_{j<0}\mathfrak{g}_{\overline{j}} \otimes
\lambda^j.
\end{gather*}
In the subsequent exposition we will also use the following
statement:

\begin{proposition}
For all the  above classical matrix Lie algebras $\mathfrak{g}$
and matrices $A$  defined in the Proposition~{\rm \ref{cond}} it
is possible to introduce the following ``pseudo-roots''
$\beta_i^A$:
\begin{gather}\label{pr}
[X,H_i]_A=\beta_i^A(X)H_i,
\end{gather}
where $X\in \mathfrak{g}_{\overline{-1}}$ and $H_i$  are  basic
elements in Cartan subalgebra defined in the Section~{\rm
\ref{general}}.
\end{proposition}

\begin{remark}
 We call linear forms $\beta_i^A$ to be
``pseudo-roots'' because  $\mathfrak{g}_A=(\mathfrak{g},[\cdot
,\cdot ])$ in the general case is not isomorphic to $\mathfrak{g}$
and $\mathfrak{g}_{\overline{-1}}\equiv
(\mathfrak{g}_A)_{\overline{0}}$ is not commutative with respect
to the $[\cdot ,\cdot ]_A$-bracket.
\end{remark}

\section[Dual space and Lie-Poisson bracket]{Dual space and Lie--Poisson bracket}

In order to describe applications of the Lie algebras
$\widetilde{\mathfrak{g}}^{\rm pr}_{A}$ to the theory of
f\/inite-dimensional integrable Hamiltonian systems it is
necessary to def\/ine linear space $(\widetilde{\mathfrak{g}}^{\rm
pr}_{A})^*$, the corresponding Lie--Poisson bracket and the
Casimir functions of $\widetilde{\mathfrak{g}}^{\rm pr}_{A}$.

\subsection[Coadjoint representation and invariant functions
of $\widetilde{\mathfrak{g}}^{\rm pr}_{A}$]{Coadjoint
representation and invariant functions of
$\boldsymbol{\widetilde{\mathfrak{g}}^{\rm pr}_{A}}$}

 In this subsection we will
construct the dual space, coadjoint representation and its
invariants for the case of the "principal'' quasigraded Lie
algebras $\widetilde{\mathfrak{g}}^{\rm pr}_{A}$.  If $h$ is the
Coxeter number then from  properties of invariant
  form on simple Lie algebras it
follows \cite{Kac} that $(\mathfrak{g}_{\overline{i}},
\mathfrak{g}_{\overline{j}})=0$ if ${i+j}\neq 0\ \mod h$. Hence we
can def\/ine pairing between $\widetilde{\mathfrak{g}}^{\rm
pr}_{A}$ and $(\widetilde{\mathfrak{g}}^{\rm pr}_{A})^*$ in the
following way:
\begin{gather}\label{pai}
\langle X(\lambda),L (\lambda) \rangle={\rm
res}_{\lambda=0}\lambda^{-1}\, {\rm Tr}\,(X(\lambda)L(\lambda)).
\end{gather}
From this def\/inition  it follows that the  generic element
$L(\lambda)\in (\widetilde{\mathfrak{g}}^{\rm pr}_A)^*$ has the
form:
\begin{gather*}
L(\lambda)=\sum\limits_{j\in Z}\sum\limits_{\alpha=1}^{\dim
\mathfrak{g}_{\overline{j}}}l^{(j)}_{\alpha}
X^{\overline{-j}}_{\alpha}\lambda^{-j},
\end{gather*}
 where $X^{\overline{-j}}_{\alpha}$ is
an element of  basis of subspace $\mathfrak{g}_{\overline{-j}}$,
$l^{(j)}_{\alpha}$ is a coordinate function on
$(\widetilde{\mathfrak{g}}^{\rm pr}_A)^*$.
\begin{remark}
We use  the notion of inf\/inite-dimensional Lie algebras
$\widetilde{\mathfrak{g}}_A$ and $\widetilde{\mathfrak{g}}^{\rm
pr}_A$ in the algebraic sense of \cite{Kac}, i.e.\ we require that
each of its elements consists of {\it finite} linear combination
of an inf\/inite set of the natural
  elements of  basis
$\{X^{\overline{j}}_{\alpha}\lambda^{j}\}$.
 On the contrary, we let the dual space~$\widetilde{\mathfrak{g}}^*$ to be wider, i.e.\ to contain also
{\it infinite} linear combination of the elements of its  basis
$(X^{\overline{j}}_{\alpha}\lambda^{j})^*=
X^{\overline{-j}}_{\alpha}\lambda^{-j}$. Under such agreement all
subsequent consideration will be correct and consistent
(see~\cite{Skr7'}).
\end{remark}

The following proposition holds true:
\begin{proposition}\label{casim}
Let $L(\lambda)\in(\widetilde{\mathfrak{g}}^{\rm pr}_A)^*$ be the
generic element of the dual space. Then the functions
\begin{gather*}
  I_{k}^m(L(\lambda))=\frac{1}{m} \, {\rm res}_{\lambda=0}
\lambda^{-k-1}\,{\rm Tr}\,
\bigl(L(\lambda)A(\lambda)^{-1}\bigr)^m,
\end{gather*}
are invariants of the coadjoint representation of the Lie algebra
$\widetilde{\mathfrak{g}}^{\rm pr}_A$
\end{proposition}

\begin{proof}  It follows from the explicit form of the coadjoint
action which  has the following form:
\begin{gather*}
{\rm ad}_{X(\lambda)}^* \circ L(\lambda)=
A(\lambda)X(\lambda)L(\lambda)-L(\lambda)X(\lambda)A(\lambda),
\end{gather*}
where $A(\lambda)=(1-\lambda A)$, $X(\lambda),
Y(\lambda)\in\widetilde{\mathfrak{g}}^{\rm pr}_A$,
$L(\lambda)\in(\widetilde{\mathfrak{g}}^{\rm pr}_A)^*$.
\end{proof}

\begin{remark}\label{rozkl}
The matrix $A(\lambda)^{-1}\equiv (1-\lambda A)^{-1}$ in the above
Proposition has to be understood as a power series in $\lambda$ in
the neighborhood of $0$ or $\infty$:
$A(\lambda)^{-1}=(1+A\lambda+A^2 \lambda^2+\cdots)$ or
$A(\lambda)^{-1}=-(A^{-1}\lambda^{-1}+A^{-2}
\lambda^{-2}+\cdots)$. The corresponding decomposition should be
chosen in such a way that the restriction of the invariant
function $I_{k}^m(L(\lambda))$ onto the dual space
$(\widetilde{\mathfrak{g}}_A^{{\rm pr}\pm})^*$ are f\/inite
polynomials (see~\cite{Skr7'}). More precisely, when restricting
invariant functions on $(\widetilde{\mathfrak{g}}_A^{{\rm
pr}-})^*$ one has to choose the decomposition of $A(\lambda)^{-1}$
in the neighborhood of $0$, and when restricting invariant
function on $(\widetilde{\mathfrak{g}}_A^{{\rm pr}+})^*$ one has
to chose the decomposition of $A(\lambda)^{-1}$ in the
neighborhood of $\infty$.
\end{remark}

\subsection[Lie-Poisson bracket]{Lie--Poisson bracket}

Let us def\/ine Poisson structures in the space
$(\widetilde{\mathfrak{g}}^{\rm pr}_{A})^*$. Using pairing
(\ref{pai})  (described in the previous section) we can def\/ine
Lie--Poisson bracket on $P((\widetilde{\mathfrak{g}}^{\rm
pr}_{A})^*)$ in the standard way:
\begin{gather}\label{br1}
\{F_1(L(\lambda)),F_2(L(\lambda))\}=  \langle L(\lambda),[\nabla
F_1(L(\lambda)),\nabla F_2(L(\lambda))]_{A(\lambda)} \rangle,
\end{gather}
here $\nabla F_i(L(\lambda))= \sum\limits_{j\in
Z}\sum\limits_{\alpha=1}^{\dim \mathfrak{g}_{\overline{j}}}
\dfrac{\partial F_i}{\partial l_{\alpha}^{(j)}
}X_{\alpha}^{\overline{j}} \lambda^j$ and
$X_{\alpha}^{\overline{j}}$ is an  element of basis of subspace
$\mathfrak{g}_{\overline{j}}$.

 From the Proposition~\ref{casim}  and standard arguments
follows the next corollary:

\begin{corollary} The
functions $I_{k}^m (L(\lambda))$ are central (Casimir) functions
for the Lie--Poisson bracket \eqref{br1}.
\end{corollary}

Let us  calculate Poisson bracket (\ref{br1}) explicitly.
 It is
easy to show that for the coordinate functions $l_{\alpha}^{(i)}$,
$l_{\beta}^{(j)}$, where
$l_{\alpha}^{(i)}\in(\mathfrak{g}_{\overline{i}})^*$,
$l_{\beta}^{(j)}\in(\mathfrak{g}_{\overline{j}})^*$, this bracket
will have the following form:
\begin{gather*}
\{l_{\alpha}^{(i)},l_{\beta}^{(j)}\}=\sum\limits_{\gamma}C_{\alpha,\beta}^{\gamma}l_{\gamma}^{(i+j)}
-
\sum\limits_{\delta}C_{\alpha,\beta}^{\delta}(A)l_{\delta}^{(i+j+1)},
\end{gather*}
where $l_{\gamma}$ and $l_{\delta}$ are the coordinate functions
on $(\mathfrak{g}_{\overline{i+j}})^*$ and
$(\mathfrak{g}_{\overline{i+j+1}})^*$. This bracket determines the
Lie algebra structure isomorphic to $\widetilde{\mathfrak{g}}^{\rm
pr}_A$ in the space of linear functions $\{l_{\alpha}^{i}\}$ and,
hence, subspaces $((\widetilde{\mathfrak{g}}^{\rm pr}_A)^{\pm})^*$
are Poisson.

\subsection[Infinite-component  Hamiltonian systems via
$\widetilde{\mathfrak{g}}^{\rm pr}_A$]{Inf\/inite-component
Hamiltonian systems via $\boldsymbol{\widetilde{\mathfrak{g}}^{\rm
pr}_A}$}
 In this subsection we
construct Hamiltonian systems on the inf\/inite-dimensional space
$(\widetilde{\mathfrak{g}}_{A}^{\rm pr})^*$ possessing inf\/inite
number of independent, mutually commuting integrals of motion. Let
\[
L^{\mp}(\lambda) = \sum\limits_{j \in
\mathbb{Z}_{\pm}}\sum\limits_{\alpha=1}^{\dim
\mathfrak{g}_{\overline{j}}
}L_{\alpha}(\lambda)X^{\overline{-j}}_{\alpha}=\sum\limits_{j \in
\mathbb{Z}_{\pm}} \sum\limits_{\alpha=1}^{\dim
\mathfrak{g}_{\overline{j}}
}l_{\alpha}^{(j)}\lambda^{-j}X^{\overline{-j}}_{\alpha}
\]
be the generic elements of the spaces
$(\widetilde{\mathfrak{g}}_{A}^{{\rm pr}\pm})^*$.  Let us consider
the functions on $(\widetilde{\mathfrak{g}}^{\rm pr}_{A})^*$ of
the form
\begin{gather*}
I^{m\mp}_{k}(L(\lambda))\equiv  I^m_{k}(L^{\mp}(\lambda)),
\end{gather*}
 where
$\{I^m_{k}(L(\lambda))\}$ are Casimir  functions of
$\widetilde{\mathfrak{g}}^{\rm pr}_{A}$. Due to the
Remark~\ref{rozkl} functions $I^{m\mp}_{k}(L(\lambda))$  are
f\/inite polynomials on $(\widetilde{\mathfrak{g}}_{A}^{{\rm
pr}\pm})^*$. The Hamiltonian f\/lows corresponding to the
Hamiltonians $I^{m\mp}_k(L(\lambda))$ are written in a standard
way:
\begin{gather}\label{ham}
\frac{\partial L_{\alpha}(\lambda)}{\partial
t^{m\mp}_k}=\big\{L_{\alpha}(\lambda),I^{m\mp}_{k}(L(\lambda))\big\}.
\end{gather}
  The following theorem is true:
\begin{theorem}\label{stheo}
 {\rm (i)} The time flows given by the equations \eqref{ham} are
 correctly defined and commute for all times $t^{m+}_k$, $t^{n-}_l$.

 {\rm (ii)} Euler--Arnold equations \eqref{ham}
are written in the ``deformed'' Lax form:
\begin{gather}\label{dlax2}
 \frac{\partial L(\lambda)}{\partial t^{m\mp}_k}=
 A(\lambda)M^{m\pm}_{k}(\lambda)L(\lambda)-
L(\lambda)M^{m\pm}_{k}(\lambda)A(\lambda).
\end{gather}
where $M^{m\pm}_{k}(\lambda)=\nabla
I^m_{k}(L^{\mp}(\lambda))=\sum\limits_{j \in
\mathbb{Z}_{\pm}}\sum\limits_{\alpha=1}^{\dim
\mathfrak{g}_{\overline{j}}} \dfrac{\partial I^m_{k}}{\partial
l_{\alpha}^{(j)} }X^{\overline{j}}_{a}\lambda^{j}$.

{\rm (iii)} The functions $I^p_{q}(L^{\pm})$ are constant along
all times $t^{m\pm}_k$ and $t^{n\mp}_l$.
\end{theorem}

The proof of this theorem repeats the proof of the analogous
theorem for the case of ordinary loop algebras (see \cite{New} and
references therein).

\begin{remark}
Note that by  virtue of the fact that functions
$I^m_{k}(L^{\mp}(\lambda))$ are f\/inite polynomials $M$-operators
$M^{m\pm}_{k}(\lambda)=\nabla I^m_{k}(L^{\mp}(\lambda))$ belong to
$\widetilde{\mathfrak{g}}^{{\rm pr}\pm}_A$, i.e.\ are f\/inite
linear combinations of the basic elements
$X^{\overline{j}}_{a}\lambda^{j}$. This fact provides the
correctness of the def\/inition of equation~(\ref{dlax2}).
\end{remark}

\section[``Modified'' Toda field equations]{``Modif\/ied'' Toda f\/ield equations}

\subsection[Zero-curvature condition with the values in
$\widetilde{\mathfrak{g}}^{\rm pr}_A$]{Zero-curvature condition
with the values in $\boldsymbol{\widetilde{\mathfrak{g}}^{\rm
pr}_A}$}

In this subsection we will obtain zero curvature-type equations
with the values in the Lie
algeb\-ras~$\widetilde{\mathfrak{g}}^{\rm pr}_A$. The following
theorem holds true:
\begin{theorem}\label{dzc} Let the infinite-dimensional Lie algebras
 $\widetilde{\mathfrak{g}}^{\rm pr}_A$,
 $\widetilde{\mathfrak{g}}_A^{{\rm pr}\pm}$, their dual spaces
   and polynomial Hamiltonians $I_k^m(L^{\pm}(\lambda))$,
   $I^n_{s}(L^{\pm}(\lambda))$ on them be defined as in previous sections.
    Then the algebra-valued gradients of
these functions  satisfy the ``deformed'' zero-curvature
equations:
\begin{gather}\label{zce1}
\dfrac{\partial \nabla I_{k}^m(L^{\pm}(\lambda))}{\partial
t^{n\pm}_l}-\dfrac{\partial \nabla
I^n_{s}(L^{\pm}(\lambda))}{\partial t^{m\pm}_k}+[\nabla
I_{k}^m(L^{\pm}(\lambda)),\nabla
I^n_{s}(L^{\pm}(\lambda))]_{A(\lambda)}=0,
\\
\label{zce2} \dfrac{\partial \nabla
I_{k}^m(L^{\pm}(\lambda))}{\partial t^{n\mp}_l}-\dfrac{\partial
\nabla I^n_{s}(L^{\mp}(\lambda))}{\partial t^{m\pm}_k}+[\nabla
I_{k}^m(L^{\pm}(\lambda)),\nabla
I^n_{s}(L^{\mp}(\lambda))]_{A(\lambda)}=0.
\end{gather}
\end{theorem}

 {\bf Idea of the Proof.} The statement of  the theorem, i.e. validity
 of the   equations (\ref{zce1}), (\ref{zce2}), follows from the
 commutativity of the ``deformed'' Lax f\/lows constructed in the
 previous section.

\begin{remark} Using the mentioned above  realizations of
  $\widetilde{\mathfrak{g}}^{\rm pr}_A$ the
``deformed'' zero-curvature and Lax equations can be rewritten in
the standard form,   but in this case the corresponding $U$--$V$
and $L$--$M$ pairs will be more complicated and forcing us to work
with the zero-curvature and Lax equations in the ``deformed'' form
(\ref{zce1}), (\ref{zce2}).
\end{remark}

Theorem \ref{dzc}  provides us with an inf\/inite number of
$\widetilde{\mathfrak{g}}^{\rm pr}_A$-valued $U$--$V$ pairs that
satisfy zero curvature-type equations. The latter are non-linear
equations in  partial derivatives in the dynamical variables
--- matrix elements of the matrix $L(\lambda)$. In the
terminology of~\cite{TF} equations generated by the inf\/inite set
of $U$--$V$ pairs are called ``integrable in the kinematic
sense''.
 In the next subsections we will consider the simplest examples of
 such
integrable equations and their hierarchies.

\subsection[Modified Toda-field equations: general case]{Modif\/ied Toda-f\/ield equations: general case}

In this subsection we obtain concrete examples of integrable
equations satisfying the ``deformed'' zero-curvature
representations constructed in the previous subsection. The most
interesting of them will be a ``modif\/ied'' Toda f\/ield
equations. We will construct these equations for all series of
classical simple (reductive) lie algebras.

Let us at f\/irst consider a general situation. Let the
``principal quasigraded'' Lie algebra
$\widetilde{\mathfrak{g}}^{\rm pr}_A$ and its decomposition
$\widetilde{\mathfrak{g}}^{\rm pr}_A=
\widetilde{\mathfrak{g}}^{{\rm pr}+}_A+
\widetilde{\mathfrak{g}}^{{\rm {\rm pr}}-}_A $  be def\/ined as in
the previous sections. Generic elements $L^{\pm}(\lambda)\in
(\widetilde{\mathfrak{g}}^{{\rm pr}\mp}_A)^*$ of the dual spaces
have the following form:
\[
L^{+}(\lambda)=\lambda L^{(-1)} +\lambda^{2} L^{(-2)}+\cdots,
\qquad L^{-}(\lambda)=L^{(0)}+\lambda^{-1} L^{(1)} +\lambda^{-2}
L^{(2)}+\cdots,
\]
 where $L^{(k)} \in \mathfrak{g}_{\overline{-k}}$.
In particular $L^{(0)}=\sum\limits_{i=1}^{n}l^{(0)}_i H_i$,
$L^{(-1)}=\sum\limits_{\alpha_i\in \Pi \bigcup -\Theta}l^{(-1)}_i
E_{\alpha_i}$.

Now we can formulate the following theorem:

\begin{theorem}
Let $\mathfrak{g}$ be one of the classical matrix Lie algebras
$gl(n)$, $so(2n+1)$, $sp(n)$ or $so(2n)$. Let the ``deformation''
matrix
 $A$ corresponding to the ``principal'' grading of $\mathfrak{g}$
   be defined as in the Proposition~{\rm \ref{cond}}. Then:

{\rm (i)} among the  corresponding equations \eqref{zce1} there is
the equation equivalent to the following ``Volterra coupled
system'':
\begin{gather}\label{vcs}
\partial_t u_i=u_i\beta^A_{i}(v), \qquad
\partial_x v_i= v_i \alpha_i(u),
\end{gather}
where $\beta^A_{i}$ are ``pseudo-roots'' on
$\mathfrak{g}_{\overline{-1}}$ defined by \eqref{pr}  and
depending on the ``deformation'' matrix~$A$ and Lie algebra
$\mathfrak{g}$, $\alpha_i \in \Pi\bigcup \{-\Theta\}$ are linear
forms (roots) on $\mathfrak{h}$, $H_i$ is the basis in
$\mathfrak{h}$, $E_{-\alpha_i}$, $E_{\Theta}$~is the basis in
$\mathfrak{g}_{\overline{-1}}$, and
\begin{gather*}
U=u=\sum\limits_{i=1}^{\dim \mathfrak{h}} u_iH_i, \qquad
V=\lambda^{-1}v=\lambda^{-1}\left(\sum\limits_{\alpha_i\in
\Pi\bigcup -\Theta } v_iE_{-\alpha_i}\right)
\end{gather*}
is the corresponding $U$--$V$ pair.

 {\rm (ii)} Volterra coupled system \eqref{vcs} is written in the form
 of the
``modified'' Toda field equations:
\begin{gather}\label{mT}
\partial^2_{xt} \psi_i=\partial_{x} \psi_i\left(\sum\limits_{\alpha_j\in
\Pi\bigcup \{-\Theta\}}
 \beta^A_{i,j} e^{\alpha_j({\psi})}\right),
\end{gather}
where ${\psi}= \sum\limits_{i=1}^n \psi_iH_i$ and $\psi_i$ satisfy
the following differential constraint
$\partial_x\psi_1\partial_x\psi_2\cdots
\partial_x\psi_n=c$.
\end{theorem}
\begin{proof}
 To prove item (i) of the theorem it is enough to
show that among the integrals $I^m_k$, $I^n_l$ are such integrals
$I^{m_0}_{k_0}$, $I^{n_0}_{l_0}$ that
$I^{m_0}_{k_0}(L^{+}(\lambda)) \equiv I^{m_0}_{k_0}(L^{(-1)})$,
$I^{n_0}_{l_0}(L^{-}(\lambda)) \equiv I^{n_0}_{l_0}(L^{(0)})$. Let
us explicitly f\/ind these functions. Let us consider the
generating series $I^{m_0}_{k_0}(L^{+}(\lambda))$:
\[
I^{m_0}_{k_0}(L^{+}(\lambda))=I^{m_0}(A^{-1}(\lambda)L^{+}(\lambda))=I^{m_0}((1+A\lambda+A^2\lambda^2+\cdots)(
\lambda L^{(-1)} +\lambda^{2} L^{(-2)}+\cdots)).
\]
Taking into account that $I^{m_0}\!$ is a homogeneous polynomial on
$\mathfrak{g}^*$ we easily obtain that
$I^{m_0}_{m_0}\!(L^{+}(\lambda))\!$ $= I^{m_0}_{m_0}(L^{(-1)})$. Now,
for the role of   $I^{m_0}$ we have to take  any Casimir function
on $\mathfrak{g}$ that has a non-trivial restriction onto
$\mathfrak{g}_{\overline{1}}$. Such Casimir function always
exists, because the generic element of the space
$\mathfrak{g}_{\overline{1}}$ is never nilpotent.

Let us now consider the generating series
$I^{n_0}(L^{-}(\lambda))$. If we chose  the generating Casimir
function as follows: $I^{n_0}(L^{-}(\lambda))= \det A(\lambda)
\,{\rm Det}\,(L^{-}(\lambda)A(\lambda)^{-1})\equiv {\rm
Det}\,L^{-}(\lambda)$, we obtain:
\[
 I^{n_0}_0(L^{-}(\lambda))= {\rm Det}\,(L^{(0)}).
\]
 For such  the Hamiltonians  their matrix
gradients have the following form: $\nabla
I^{m_0}_{m_0}(L^{+}(\lambda))\equiv u=\sum\limits_{i=1}^n u_iH_i$,
$\nabla I^{n_0}_{0}(L^{-}(\lambda))\equiv
\lambda^{-1}v=\lambda^{-1}(\sum\limits_{\alpha_i\in \Pi}
v_iE_{-\alpha_i}+v_{0}E_{\Theta})$, where $u_i\equiv \frac{\partial I^{m_0}_{m_0} }{\partial
l^{(0)}_i}$, $v_i\equiv \frac{\partial I^{n_0}_{0} }{\partial
l^{(-1)}_i}$. Substituting
this into the corresponding zero-curvature condition (\ref{zce2})
we obtain item (i) of the theorem.

   Making the substitution
of variables: $u_i=\partial_x \psi_i$, $v_i=e^{\alpha_i(\psi)}$ we
obtain equation (\ref{mT}). Finally, taking into account the
constancy of the Hamiltonian $I^{n_0}_{l_0}(L^{(0)})$ along all
time f\/lows  we obtain that $\prod\limits_{i=1}^n u_i={\rm
const}$. That proves item (ii).
\end{proof}

\begin{remark} In  \cite{ShYa,LesSav} $gl(n)$
coupled Volterra systems were obtained using the B\"acklund
transformation for the Toda integrable hierarchy. Such connection
means that starting from the known integrable hierarchy and
B\"acklund transformation one can  generate a new integrable
hierarchy (see \cite{BZ}). The investigation of the possibility of
such connection for the Volterra coupled systems, associated with
the other series of classical matrix Lie algebras, is a separate
problem and we will return to it in our subsequent publications.
\end{remark}

\subsection[Example: $gl(n)$-modified Toda field equation]{Example: $\boldsymbol{gl(n)}$-modif\/ied Toda f\/ield equation}

Let us consider the Lie algebra $\widetilde{gl(n)^{\rm pr}_A}$
def\/ined in the Example 1 and its natural decomposition into the
direct sum $\widetilde{gl(n)^{\rm pr}_A}= \widetilde{gl(n)^{{\rm
pr}+}_A}+ \widetilde{gl(n)^{{\rm pr}-}_A} $ where
$\widetilde{gl(n)^{{\rm pr}+}_A}= \sum\limits_{j\geq 0}
gl(n)_{\overline{j}} \lambda^{j}$, $\widetilde{gl(n)^{{\rm
pr}-}_A}= \sum\limits_{j<0} gl(n)_{\overline{j}} \lambda^{j}$ and
$\overline{j}=j\mod n$. The generic elements $L^{\pm}(\lambda)\in
(\widetilde{gl(n)^{{\rm pr}\mp}_A})^*$ of the dual spaces have the
following form:
\[
L^{+}(\lambda)=\lambda L^{(-1)} +\lambda^{2} L^{(-2)}+\cdots,
\qquad L^{-}(\lambda)=L^{(0)}+\lambda^{-1} L^{(1)} +\lambda^{-2}
L^{(2)}+\cdots,
\]
 where $L^{(k)} \in \mathfrak{g}_{\overline{-k}}$:
$L^{(0)}=\sum\limits_{i=1}^n l_i^{(0)} X_{ii}$,
$L^{(-1)}=\sum\limits_{i=1}^{n-1}l_{i}^{(-1)} X_{i, i+1} +
l^{(-1)}_nX_{n1}$. The simplest  integrals are:
\begin{gather*}
D^{n-}_0( L(\lambda))=  l^{(0)}_{1} l^{(0)}_{2}\cdots
l^{(0)}_{n},\qquad D^{n+}_n( L(\lambda))=l_{1}^{(-1)} l_{2}^{(-1)}
\cdots l_{n-1}^{(-1)}l_{n}^{(-1)}.
\end{gather*}
Their matrix gradients yield the following $U$--$V$ pair:
\begin{gather*}
U\equiv\nabla D^{n-}_0=\sum\limits_{i=1}^n u_i X_{ii},\qquad V
\equiv \nabla
D^{n+}_n=\lambda^{-1}\left(\sum\limits_{i=1}^{n-1}v_{i}X_{i+1 i} +
v_nX_{1n}\right).
\end{gather*}
where $u_i\equiv D^{n-}_0/l_i^{(0)}$, $v_i\equiv
D^{n+}_n/l_i^{(-1)}$.

The corresponding ``deformed'' zero-curvature condition:
\[
\dfrac{\partial \nabla D_{0}^{n-}(L(\lambda))}{\partial
t}-\dfrac{\partial \nabla D^{n+}_{n}(L(\lambda))}{\partial
x}+[\nabla D_{0}^{n-}(L(\lambda)),\nabla
D^{n+}_{n}(L(\lambda))]_{A(\lambda)}=0,
\]
yields the following equations\footnote{In the case $a_i\neq 0$
one may make   $a_i=1$ rescaling  the variables $v_i$.}:
\begin{alignat*}{3}
& \partial_t u_1=u_1(a_1v_1-a_nv_n), \qquad &&  \partial_x
v_1=v_1(u_2-u_1),&
\\
& \partial_t u_2=u_2(a_2v_2-a_1v_1), \qquad &&  \partial_x
v_2=v_2(u_3-u_2), &
\\
& \cdots  \cdots    \cdots      \cdots    \cdots     \cdots
\cdots     \cdots &&
 \cdots  \cdots    \cdots      \cdots      \cdots     \cdots      & \\
& \partial_t u_n=u_n(a_{n}v_n-a_{n-1}v_{n-1}),\qquad&&  \partial_x
v_n=v_n(u_1- u_n).&
\end{alignat*}
This system of the hyperbolic equations is the so-called
``Volterra coupled systems''. In the case of inf\/inite $n$ and
$a_i\equiv 1$ it was f\/irst considered in the paper~\cite{ShYa}.
In the non-periodic case it was also considered in the
book~\cite{LesSav}.

This system goes together with the two constraints:
\[
u_1u_2\cdots u_n={\rm const}_1,\qquad v_1v_2\cdots v_{n-1}v_n={\rm
const}_2,
\]
They follow from the constancy of the Hamiltonians $D_{0}^{n-}$
and $D^{n+}_{n}$   with respect to the all time f\/lows.

 In order for the
``Volterra coupled systems'' to acquire the form of the periodic
``modif\/ied'' Toda system  it is enough to make the following
change of variables~\cite{ShYa}:
\[
u_i=\partial_x\psi_i, \quad i=1,n, \qquad
v_i=e^{\psi_{i+1}-\psi_{i}}, \quad i=1,n-1,\qquad v_n=
e^{\psi_{1}-\psi_{n}}.
\]
  in the result we  obtain the
following system of equations:
\begin{gather*}
\partial^2_{xt}
\psi_1=\partial_x\psi_1(a_1e^{\psi_{2}-\psi_{1}}-a_ne^{\psi_{1}-\psi_{n}}), \\
\partial^2_{xt}
\psi_2=\partial_x\psi_2(a_2\epsilon^{\psi_{3}-\psi_{2}}-a_1e^{\psi_{2}-\psi_{1}}),  \\
\cdots    \cdots           \cdots \cdots    \cdots
\cdots \cdots  \cdots           \cdots
  \cdots          \cdots     \cdots     \\
\partial^2_{xt} \psi_n=\partial_x\psi_n(a_{n}e^{\psi_{1}-\psi_{n}}- a_{n-1}e^{\psi_{n}-\psi_{n-1}})
\end{gather*}
with additional dif\/ferential constraint:
\begin{gather*}
\partial_x\psi_1\partial_x\psi_2\cdots \partial_x\psi_n=c.
\end{gather*}

The simplest case, when the corresponding constraint can be
solved, is the case $\partial_x\psi_i=0$ for some f\/ixed $i$.
This is equivalent to the reduction $u_i=0$ in the corresponding
Volterra coupled system. Let us put, for example, $\psi_{n}={\rm
const}=0$. In this case we obtain the following ``open''
$(n-1)$-component modif\/ied Toda chain:
\begin{gather*}
\partial^2_{xt} \psi_1=\partial_x\psi_1(a_1e^{\psi_{2}-\psi_{1}}-
a_ne^{\psi_{1}}), \\
\partial^2_{xt}
\psi_2=\partial_x\psi_2(a_2\epsilon^{\psi_{3}-\psi_{2}}-a_1e^{\psi_{2}-\psi_{1}}),  \\
\cdots  \cdots    \cdots   \cdots \cdots  \cdots    \cdots   \cdots   \cdots  \cdots    \cdots   \cdots                  \\
\partial^2_{xt} \psi_{n-1}=\partial_x\psi_{n-1}(a_{n-1}e^{-\psi_{n-1}}-
a_{n-2}e^{\psi_{n-1}-\psi_{n-2}}).
\end{gather*}

\subsection[Example: $\mathfrak{g}=so(2n)$-modified Toda field equation]{Example:
 $\boldsymbol{\mathfrak{g}=so(2n)}$-modif\/ied Toda f\/ield equation}

Let us consider the Lie algebra $\widetilde{so(2n)^{\rm pr}_A}$
being def\/ined in the Example~4 and its natural decomposition
into the direct sum $\widetilde{so(2n)^{\rm pr}_A}=
\widetilde{so(2n)^{{\rm pr}+}_A}+ \widetilde{so(2n)^{{\rm
pr}-}_A}$.
 Generic elements
$L^{\pm}(\lambda)\in (\widetilde{so(2n)^{{\rm pr}\mp}_A})^*$ of
the dual spaces have the following form:
\[
L^{+}(\lambda)=\lambda L^{(-1)} +\lambda^{2} L^{(-2)}+\cdots,
\qquad L^{-}(\lambda)=L^{(0)}+\lambda^{-1} L^{(1)} +\lambda^{-2}
L^{(2)}+\cdots,
\]
 where $L^{(k)} \in so(2n)_{-k \mod
(2n-2)}$.
\begin{gather*}
L^{(0)}=\sum\limits_{i=1}^{n}l^{(0)}_i (X_{ii}-X_{i+n i+n}),\\
L^{(-1)}=\sum\limits_{i=1}^{n-1}l^{(-1)}_i (X_{i i+1}-X_{i+n+1
i+n})+ l^{(-1)}_n (X_{n 2n-1}-X_{n-1 2n})+ l^{(-1)}_{n+1}
(X_{n+12}-X_{n+21}).
\end{gather*}
It turned out that the form of the needed Casimir functions and
Hamiltonians dif\/fers for the cases $n\leq 3$ and $n>3$. We will
consider here only the case $n>3$.
 The simplest non-trivial integrals are:
\begin{gather*}
D^{2n-}_0( L(\lambda))=  \big(l^{(0)}_{1} l^{(0)}_{2}\cdots
l^{(0)}_{n}\big)^2,\\
T^{(2n-2)+}_{2n-2}( L(\lambda))=l_{1}^{(-1)}
\big(l_{2}^{(-1)}\big)^2 \cdots
\big(l_{n-2}^{(-1)}\big)^2l_{n-1}^{(-1)}
l_{n}^{(-1)}l^{(-1)}_{n+1}.
\end{gather*}
{\samepage Their matrix gradients are:
\begin{gather*}
U\equiv\nabla D^{2n-}_0= \sum\limits_{i=1}^{n}u_i
(X_{ii}-X_{i+n i+n}),\\
 V\equiv\nabla T^{(2n-2)+}_{2n-2}=\lambda^{-1}
 \Bigg (\sum\limits_{i=1}^{n-1}v_i (X_{i+1 i}-X_{i+n i+n+1})+ v_n
(X_{2n-1 n}-X_{2n n-1})\\
\phantom{ V\equiv\nabla T^{(2n-2)+}_{2n-2}=}{}+ v_{n+1} (X_{2
n+1}-X_{1 n+2})\Bigg),
\end{gather*}
where $u_i\equiv D^{2n-}_0/l_i^{(0)}$, $v_i \equiv
T^{(2n-2)+}_{2n-2}/l_i^{(-1)}$.}

The corresponding zero-curvature condition
\begin{equation*}
\dfrac{\partial \nabla D_{0}^{2n-}(L(\lambda))}{\partial
t}-\dfrac{\partial \nabla T^{(2n-2)+}_{2n-2}(L(\lambda))}{\partial
x}+\big[\nabla
D_{0}^{2n-}(L(\lambda)),T^{(2n-2)+}_{2n-2}(L(\lambda))\big]_{A(\lambda)}=0,
\end{equation*}
yields the following equations:
\begin{alignat*}{3}
&\partial_t u_1=u_1(a_1v_1-a_{n+1}v_{n+1}),&&   \partial_x v_1=v_1(u_2-u_1), & \\
&\partial_t u_2=u_2(a_2v_2-a_1v_1+a_{n+1}v_{n+1}), & & \partial_x v_2=v_2(u_3-u_2), &  \\
& \partial_t u_3=u_3(a_3v_3-a_2v_2), & & \partial_x v_3=v_3(u_4-u_3),&\\
& \cdots \cdots\cdots\cdots\cdots\cdots\cdots\cdots\cdots\cdots\cdots\cdots & & \cdots\cdots\cdots\cdots\cdots\cdots \cdots &  \\
&    \partial_t u_{n-2}=u_{n-2}(a_{n-2}v_{n-2}-a_{n-3}v_{n-3}), & &\partial_x v_{n-1}=v_{n-1}(u_n-u_{n-1}), &    \\
& \partial_tu_{n-1}=u_{n-1}(a_{n-1}v_{n-1}-a_{n-2}v_{n-2}-a_nv_n),\qquad && \partial_x v_{n}=-v_{n}(u_n+u_{n-1}), &    \\
& \partial_t u_n=u_n(a_{n}v_n-a_{n-1}v_{n-1}), &&  \partial_x
v_{n+1}=v_{n+1}(u_1+u_2).&
\end{alignat*}
This system goes together with the natural constraints:
\[
u_1u_2\cdots u_n={\rm const}_1,\qquad v_1(v_2)^2\cdots
(v_{n-2})^2v_{n-1} v_nv_{n+1}={\rm const}_2.
\]
 which follows from the constancy of the
Hamiltonians $D_{0}^{2n-}$ and $T^{(2n-2)+}_{2n-2}$ with respect
to the all time f\/lows. Making the following replacement of
variables:
\begin{gather*}
u_i=\partial_x\psi_i, \qquad v_i=e^{\psi_{i+1}-\psi_{i}}, \quad
{\rm where}\quad i=1,n-1,\\ v_n=e^{-(\psi_{n+1} + \psi_n)}, \qquad
v_{n+1}= e^{-(\psi_{2} + \psi_1)}
\end{gather*}
 we obtain the {\it $so(2n)$-modified Toda chain}:
\begin{gather*}
\partial^2_{xt} \psi_1 =\partial_x\psi_1(a_1e^{\psi_{2}-\psi_{1}}-
a_{n+1} e^{\psi_{2} + \psi_1}), \\
\partial^2_{xt}
\psi_2=\partial_x\psi_2(a_2e^{\psi_{3}-\psi_{2}}-
a_1e^{\psi_{2}-\psi_{1}}+a_{n+1}e^{\psi_{2} +
\psi_1}),\\
\partial^2_{xt} \psi_3=\partial_x\psi_3(a_3e^{\psi_{4}-\psi_{3}}-a_2
e^{\psi_{3}-\psi_{2}}),\\
\cdots\cdots\cdots\cdots\cdots\cdots\cdots\cdots\cdots\cdots \cdots   \cdots   \\
    \partial^2_{xt} \psi_{n-2}=\partial_x\psi_{n-2}
    (a_{n-2}e^{\psi_{n-1}-\psi_{n-2}}-a_{n-3}e^{\psi_{n-2}-\psi_{n-3}}),
\\
    \partial^2_{xt} \psi_{n-1}=\partial_x\psi_{n-1}(a_{n-1}
    e^{\psi_{n}-\psi_{n-1}}
    -a_{n-2}e^{\psi_{n-1}-\psi_{n-2}}-a_ne^{-(\psi_{n-1}+\psi_{n})}),
    \\
\partial^2_{xt}
\psi_n=\partial_x\psi_n(a_{n}e^{-(\psi_{n-1}+\psi_{n})}- a_{n-1}
e^{\psi_{n}-\psi_{n-1}}).
\end{gather*}
These equations goes also with additional dif\/ferential
constraint: $\partial_x\psi_1\partial_x\psi_2\cdots
\partial_x\psi_n=c$.

\section{Conclusion and discussion}
In the present paper we have constructed  a  family of quasigraded
Lie algebras that coincide with  deformations of the ``principal''
subalgebras of  the loop algebras  and admit Kostant--Adler--Symes
scheme.
  Using the constructed
  algebras we have obtained new Volterra
 coupled systems and  modif\/ied Toda f\/ield equations for all series of
  classical matrix Lie algebras $\mathfrak{g}$.

Further, more detailed  investigation of the constructed
integrable hierarchies would be very interesting. In particular,
an interesting open problem is to f\/ind a possible connection of
the constructed new Volterra coupled systems and new modif\/ied
Toda f\/ield equations with the B\"acklund transformations and the
usual Toda f\/ield equations \cite{Mikh1,DS,LesSav}. It would be
also  interesting to f\/ind explicit forms of the simplest
equations of the ``negative'' and ``positive'' subhierarchies of
the constructed integrable hierarchies in the case of
general~$\mathfrak{g}$. They
 should coincide with the ``deformations''
of the generalized mKdV equations  \cite{dGHM} (see~\cite{Skr8}
where $gl(2)$ case was considered in details). It would be also
very nice to f\/ind out whether there exists an analog of the
``Drienf\/ield--Sokolov'' reduction \cite{DS}  and the
corresponding ``deformed'' KdV equations.

Another interesting development is to generalize the construction
of the present paper onto the case of the other graded subalgebras
of the loop algebras (see \cite{Kac,dGHM}). The work in this
direction is now in progress and the corresponding results will
soon be published.

\subsection*{Acknowledgements}
Author is grateful to P.~Holod, S.~Zykov and M.~Pavlov  for
discussions. The research was supported by INTAS YS Fellowship
Nr~03-55-2233 and  by the French-Ukrainian project ``Dnipro''.

\LastPageEnding

\end{document}